\documentclass[12pt]{article}
\usepackage{epsfig, amssymb, graphicx, amsmath} 

\pdfoutput=1

\usepackage[english]{babel}
\usepackage{epstopdf}	
\usepackage{booktabs}   	
\usepackage{eurosym}	
\usepackage{wrapfig}	
\usepackage{float}		
\usepackage{footnote}
\usepackage{amsthm} 	 

\usepackage{bbm}		

\usepackage[authoryear]{natbib}
\bibpunct{(}{)}{,}{a}{}{,}

\numberwithin{theorem}{section} 

\theoremstyle{definition}

\theoremstyle{remark}

\newcommand{\be}{\begin{equation}}
\newcommand{\en}{\end{equation}}
\newcommand{\ben}{\begin{equation*}}
\newcommand{\enn}{\end{equation*}}
\newcommand{\bea}{\begin{eqnarray}}
\newcommand{\ena}{\end{eqnarray}}

\begin{document}
 
\newlength\tindent
\setlength{\tindent}{\parindent}
\setlength{\parindent}{0pt}
\renewcommand{\indent}{\hspace*{\tindent}}

\begin{savenotes}
\title{
\bf{The measure of model risk \\ in credit capital requirements}}
\author{
Roberto Baviera
}

\maketitle

\vspace*{0.11truein}
\begin{tabular}{ll}
$ $ &  Politecnico di Milano, Department of Mathematics, 32 p.zza L. da Vinci, 20133 Milano \\
\end{tabular}
\end{savenotes}

\vspace*{0.11truein}

\begin{abstract}
\noindent
Credit capital requirements in Internal Rating Based approaches require the calibration of two key parameters:
the probability of default and the loss-given-default.
This letter considers the uncertainty about these two parameters 
 and models this uncertainty in an elementary way: it shows how
this estimation 
risk can be computed and properly taken into account in regulatory capital.

We analyse two standard real datasets: one composed by all corporates rated by Moody's and one limited only to the
speculative grade ones.  
We statistically test model hypotheses on both marginal distributions and parameter dependency. 
We compute the estimation risk impact and observe that
parameter dependency raises substantially the tail risk  in capital requirements. 
The results are striking with a required increase in regulatory capital in the range $38\%$-$66\%$.

\end{abstract}

\vspace*{0.11truein}
{\bf Keywords}: 
Regulatory capital, estimation risk, VaR, IRB approach, LGD-PD dependency.
\vspace*{0.11truein}

{\bf JEL Classification}: 
C51, 
G21, 
G28, 
G32. 

\vspace{0.6cm}
\begin{flushleft}
{\bf Address for correspondence:}\\
Roberto Baviera\\
Department of Mathematics \\
Politecnico di Milano\\
32 p.zza Leonardo da Vinci \\ 
I-20133 Milano, Italy \\
Tel. +39-02-2399 4575\\
roberto.baviera@polimi.it
\end{flushleft}

\newpage

\begin{center}
\Large\bfseries 
The measure
of model risk \\ in credit capital requirements
\end{center}


\vspace*{0.21truein}


\section{Introduction}
\label{sec:Introduction}

Following the Great Financial crisis of 2007, model risk in the capital requirement of financial institutions has emerged as a key public concern.
The set of international banking rules  require banks to hold a minimum capital as a buffer against future loss exposures.
The variability of these future losses 
depends, on the one hand, on a model via some stochastic risk factors and, 
on the other hand, on the uncertainty related to the 
selected model.
In this letter we focus on the model risk associated to credit  regulatory capital.

\smallskip 

Credit capital requirement (also known as capital adequacy) in banks is often determined via an Internal Rating Based (IRB) approach under the Basel II capital accords and the modifications 
that have followed.\footnote{The other approach, the Standardized one, presents no model risk.}
A peculiarity of the IRB approach is that the modeling framework is established by the regulator while model calibration is left to the banks.
In the IRB approach, regulators base the
capital requirement on the value-at-risk (VaR) of bank's credit portfolio
calculated
using the Asymptotic Single Risk Factor (ASRF) model, introduced by \citet{Gordy}. 
Thus, model risk  in credit capital requirements consists in the estimation noise of the parameters,
i.e. the risk arising from errors in model parameters when we cannot rely on the assumption that the parameters of the model are known with certainty.
The characteristics of credit exposures are captured by two parameters for each obligor in bank's portfolio:
the obligor's probability of default (PD)
and his loss-given-default (LGD).
The other model parameter, the correlation between obligors' assets, is established in Basel requirements for IRB
as a deterministic function of the default probability \citep[cf.][p.13]{Basel2005}.
Both parameters (PD and LGD) are the corresponding forecast over a one-year time horizon; 
they are calibrated with Through-the-Cycle values, i.e. long term default and recovery rates,  often provided by rating 
agencies.\footnote{More precisely two are the IRB approaches (Foundation and Advanced). 
Under the Foundation IRB approach, banks supply their own estimates of PD, while the other parameters are supervisory values set by the Basel Committee. 
Under the Advanced IRB approach, banks supply both PD and LGD \citep[see e.g.][]{Hull_RM, Wernz}.}

\smallskip 
Should  capital requirement  for credit risk account for parameter uncertainty? 
Albeit the relevance of this question is well known, there is a relative paucity of empirical studies that measure model risk in credit capital requirements.
The problem has been introduced by \citet{Loffler2003}, even before the details of the IRB approach were introduced by the Basel Committee:
he has analysed the impact on the $\alpha$-quantile of two homogeneous reference portfolios, one rated BBB and another one B.
After the seminal paper of  \citet{Loffler2003}, the main contribution is due to 
\cite{Tarashev2010}, who has formalized the 
model risk approach in the ASRF,
clarifying that 
the {\it correct} capital requirement reflects all potential losses, whose uncertainty includes
the imperfect information about risk parameters.

In both studies, the impact of parameter uncertainty on measures of tail risk is
analysed on the basis of a stylized credit portfolio that is homogeneous, 
 i.e. characterized by the same exposure and the same parameters (PD and LGD)  for each obligor. We follow the same approach in this letter.

We assume some probability density functions (hereinafter p.d.f.) of model parameters and statistically test these distributional hypotheses on a real dataset.
Having the parameters' p.d.f., it is possible to evaluate how much their uncertainty impacts capital requirements. 

\smallskip 

In \citet{Loffler2003} $PD$ and $LGD$ were considered independent, while LGD  was considered a constant parameter in \citet{Tarashev2010}.
However, it quite reasonable to observe a relationship between PD and LGD (or equivalently recovery). 
The economic reason is rather simple: 
if an economy experiences a recession,
on the one hand, 
the observed frequency of corporate defaults increases
and, 
on the other hand, recoveries decrease because 
the assets of failed companies are sold 
when many other firms have defaulted and
when few buyers are available
at extremely discounted prices (fire sale).\footnote{
A dependency between PD and LGD has been first pointed out by \citet{Frye2000Depressing} for non-financial issuers 
domiciled in the USA in the time interval 1982-1997,
then a positive correlation between PD and LGD has been identified and measured by \citet{Altman2005} in the speculative grade USA bond market.}
In this letter, we consider this dependency, estimate statistically it and identify the impact on capital requirement:
we show that this dependency is the most relevant source of model risk in credit capital requirements.

\smallskip 

The contributions of this letter to the existing literature are threefold:
i) we indicate some distributional assumptions for model parameters in capital requirements and statistically test them on a real dataset,
ii) we consider model risk for credit capital requirements within the IRB approach and analyse the impact of parameter dependency in capital requirements, and
iii) we draw some policy implications for credit capital requirements.



\bigskip

The rest of the letter is divided as follows. 
In Section \ref{sec:Model} we briefly recall the credit regulatory capital, describing  a {\it na\"ive} approximation and the {\it correct} implementation of the requirements in
the IRB approach.  
In Section \ref{sec:Dataset} we describe the estimation risk methodology:
we introduce a distributional assumption for model parameters and statistically test it.
In Section \ref{sec:Results} we discuss the consequences on capital requirements. 
Section \ref{sec:Conclusions} concludes focusing on  policy implications.
Finally, at the end of this letter, we recall the notation and the abbreviations we use.

\section{The capital requirement in the IRB approach}
\label{sec:Model}

In the IRB approach, the capital requirement is the 1-year unexpected loss VaR
at the $\alpha$ confidence level,
i.e. it is 
the maximum portfolio unexpected loss that is
exceeded within a year with probability $\alpha$ ($\alpha$-quantile).
The capital adequacy per unit exposure at default  (hereinafter regulatory capital
or RC) is
\[
RC = VaR_\alpha[L] - \mathbb{E}[L] \; ,
\]
where $L$ is the {\it portfolio loss rate}, i.e. the ratio of total losses to total portfolio exposure at 
default.\footnote{In this letter the exposure at 
default is not considered a source of model risk. Furthermore, IRB maturity adjustment is assumed equal to one.
}   
The value of $\alpha$ is equal to $99.9\%$  
as established by the Basel Committee for credit risk \citep[cf.][p.11]{Basel2005}.

\smallskip

Bank capital rules are 
based on the ASRF model \citep{Gordy}. 
In this section we briefly recall the main modeling results; the notation follows closely the one in  \citet{Tarashev2010}. 

We focus on model risk for a homogeneous portfolio as in \citet{Loffler2003} and \citet{Tarashev2010}.
The ASRF model, applied to a homogeneous portfolio of $n$ obligors, describes
log-assets of a generic obligor $i$ as
\ben
\begin{split}
&X_i = \sqrt{\rho} \, M +\sqrt{1-\rho} \, \varepsilon_i \quad ~i=1,\dots,n \\	&\text{and }M,~\varepsilon_i~\mathrm{are ~i.i.d.~ st.n. \, rvs.} 
\end{split}
\enn
The variable $M$ is a common credit risk factor representing the market, $\varepsilon_i$ is an obligor-specific risk component, 
while $\rho$ is the correlation between obligors' assets.  A default of an obligor in one year occurs with a probability $PD$ and produces a loss-given-default $LGD$. 
Obligor $i$ defaults when $X_i$ is below some threshold $k$, that is often referred to
as the default point \citep[see, e.g.][p.2066]{Tarashev2010}.
The threshold $k$ is then chosen s.t. $\mathbb{P}(X_i<k)=PD$; this implies $k=\Phi^{-1}(PD)$, 
with $\Phi$ the  standard normal (st.n.) cumulative distribution function. 

\smallskip

The IRB approach considers the case with a large number of obligors $n$:
the financial literature refers to this limiting portfolio as an ``asymptotic portfolio".
In this case,  the expected loss conditional on the common risk factor $M$ and given the parameters $PD$ and $LGD$, is
\be
\mathbb{E}[L| M, PD, LGD] = LGD \cdot \Phi \left(\frac{k-\sqrt{\rho} \, M}{\sqrt{1-{\rho}}}\right) \; .
\label{eq:expectedLoss}
\en
 Moreover, 
 $\rho$ is a deterministic function of $PD$ as established by the Basel Committee for IRB models \citep[cf.][p.13]{Basel2005}.
In the case of corporate, sovereign, and bank exposures this function is
\begin{equation}
\rho(PD)=0.12\cdot\frac{1-e^{-50\cdot PD}}{1-e^{-50}} +0.24\cdot\left(1-\frac{1-e^{-50\cdot PD}}{1-e^{-50}}\right) \; ,
\label{eq:rho Basel}
\end{equation}
and a similar relation holds in the other cases.

\bigskip

A {\it na\"ive} approximation of IRB  \citep[see][]{Tarashev2010} accounts for the credit risk factor $M$ 
but treats the PD and the LGD as known and equal to $\hat{PD}$ and $\hat{LGD}$, the 
point estimates of the respective parameters.
In this special case, the above setup  reduces to a single risk factor model and
the capital requirement becomes
\be
RC^{naive} = \hat{LGD} \cdot \Phi \left(\frac{ \Phi^{-1}(\hat{PD}) - \sqrt{\rho(\hat{PD})} \, \Phi^{-1}(1- \alpha)}{\sqrt{1-{\rho}(\hat{PD}) }}\right) - EL^{naive}  
\label{eqn:naiveRC}
\en
where 
\[
EL^{naive} := \hat{LGD} \cdot  \hat{PD} \; 
\]
is the expected loss and $\rho(PD)$ is given by \eqref{eq:rho Basel}. 
As already discussed by \citet{Tarashev2010}, 
the simplicity of this analytical closed formula 
justifies the popularity of 
this {\it na\"ive} IRB approach
among  practitioners.

\bigskip

In general, parameters could carry a significant estimation noise that cannot be neglected. 
Thus, the {\it correct} capital requirement in IRB is
\be
RC = VaR_\alpha \left[ \mathbb{E}[L| M, PD, LGD] \right] - \mathbb{E}[L] \; ,
\label{eq:RC}
\en
where $ \mathbb{E}[L| M, PD, LGD] $ is reported in \eqref{eq:expectedLoss}.
This result can be proven 
as in 
\citet[][proposition 1, p.2067]{Tarashev2010}.

Moreover, as already  pointed out by \citet{Tarashev2010}, $M$ is independent from parameters.
Since the
uncertainty about $M$ refers to the {\it ex-post}
realization of the credit risk factor, 
while parameters are the best {\it ex-ante} estimation given past data,
the assumed temporal independence of the risk factor $M$ implies that it is independent from parameter  uncertainty.

Hence,
there is no closed formula for the  {\it correct} capital requirement \eqref{eq:RC} but it can be easily obtained via
a Monte Carlo simulation.
It requires i) to simulate $N_{sim}$ values for the unknown parameters and the common risk factor, 
ii) to compute the loss for each simulation and
iii) to determine the $\alpha$-quantile and the mean of the loss distribution. 
Let us mention that the expected loss $\mathbb{E}[L]$ is equal to $EL^{naive}$ if the two parameters are independent, but this result does not hold true in general.

\bigskip

The aim of this research
is to consider a p.d.f. for the set of parameters in the Basel IRB approach, to statistically test the distributional assumptions and 
to evaluate the impact of the uncertainty of each parameter on the capital requirement and in particular the impact of PD-LGD dependency.

In this letter, we model the default point $k$ and the loss-given-default as Gaussian rvs; a distributional assumption that can be easily tested on a real dataset.
We consider  
\ben
\left\{
\begin{array}{lcl}
	k & \sim  & \mathbb{N}(\hat{k}, \sigma^2_k) \\[2mm]
	LGD & \sim & \mathbb{N}(\hat{LGD}, \sigma^2_{LGD}) 
\end{array}
\right. \; .
\enn
As already mentioned,
$\hat{PD} :=  \mathbb{E}[PD]$ and $\hat{LGD} :=  \mathbb{E}[LGD]$, 
while $\sigma^2_k$ and $\sigma^2_{LGD}$ are respectively the variance of $k$ and of LGD. 
The value of $\hat{k}$ can be obtained  inverting 
\be
\hat{PD}=\mathbb{E}[\Phi(k)] = \int^\infty_{-\infty} dx \, \displaystyle \frac{e^{-x^2/2}}{\sqrt{2 \pi}} \, \Phi(\hat{k} + \sigma_k \, x)
\simeq  \Phi(\hat{k})-\frac{\sigma^2_k}{2}\cdot\frac{\hat{k}}{\sqrt{2\pi}} \, e^{-\hat{k}^2/2} 
\, ,
\label{eq:kHat}
\en
where the right term is obtained via a Taylor expansion in $\sigma_k$ up to the third order. 

\bigskip
What really matters is the increase in capital requirement w.r.t. the {\it na\"ive} IRB approach.

As already mentioned,
the regulatory capital refers only to the amount of capital an institution must hold against unexpected loss.
Regulators recognize that expected losses are usually covered by the way a financial institution prices its products;
thus, if the bank computes its RC according to the {\it na\"ive} IRB, only $EL^{naive}$ are the expected losses covered by reserves when pricing products.

In presence of estimation noise, the institution must hold an {\it excess loss reserve} to cover for an increase in
both the unexpected and the expected loss.
This {\it excess loss reserve} is the additional required capital in the case a bank has considered 
a {\it na\"ive} approach; it is the difference between the VaR in the case with estimation noise and the one in the {\it na\"ive} IRB.
In particular, the relevant quantity is the (regulatory capital) {\it add-on}. 
It is defined as the ratio between the {\it excess loss reserve} (inclusive of the expected loss correction) 
and the RC in the {\it na\"ive} approximation
\be
{\textit{add-on}} := \frac{(RC - RC^{naive}) + ( \mathbb{E}[L] - EL^{naive}) }{RC^{naive}} \; .
\label{eq:addon}
\en

In this letter 
we focus on this percentage  increase in $RC^{naive}$ induced by parameter uncertainty:
we first consider the {\it  add-on} generated by each parameter one at a time and
then we analyse  all parameters together.

\section{The dataset}
\label{sec:Dataset} 

We analyse a dataset provided by Moody's Investor Service
on annual LGD rates for defaulted senior unsecured corporate bonds and on annual corporate default rates \citep[]
[exhibit 29 
\&  exhibit 41] 
{Moodys2020}.\footnote{LGD rates are obtained via Trading prices recoveries. 
Trading prices recoveries are the trading prices of defaulted 
bonds in the distressed debt market shortly after the default event: 
Moody's Investor Service reports them as the prices at which they trade about 30 days after default, 
as a percent of their face value. 
Only for distressed exchanges, Trading prices recoveries correspond to the exchange value at the default date.} 
Two are the default rates considered: the first set includes all corporates rated by Moody's (hereinafter ``All Ratings" or ``AR") while the second is limited only to firms 
who have a speculative grade 
at the beginning of the default year (hereinafter ``Speculative Grade" or ``SG"). 
The dataset reports an annual value for 
the period 1983-2019 ($37$ years); it is used by several financial institutions 
either in the determination of regulatory capital 
or in the definition of benchmarks for measuring IRB parameters.

\begin{table} [h!]                                             
		\centering                                               
\begin{tabular}{|c|c|c|c|c|c|}                              
			\hline                                                        
			& min & max & mean & median & std \\                       
			\hline \hline                 
			LGD  &  36.25\%	& 78.81\%	& 55.26\%	& 54.76\% & 10.25\% \\        
			\hline                                                      
			$PD_{AR}$  & 0.35\%	& 5.00\%	& 1.59\% &	1.25\% &	1.01\% \\     
			\hline         
			$PD_{SG}$  & 0.94\% &	12.09\%	& 4.30\% &	3.54\% &	2.62\% \\
			\hline                                                                                                        
		\end{tabular}                                     
\caption{\small Descriptive statistics for annual loss-given-default rates for senior unsecured corporate bond (LGD),
annual corporate default rates for all rated firms ($PD_{AR}$) and   
for speculative grade firms  ($PD_{SG}$).  
The data are collected at world level in the time window 1983-2019 ($37$ years).
We report min, max, mean, median and standard deviation (std).}                                    
\label{table:DescrStats}                                  
\end{table}

In this letter, we estimate the empirical properties of one-year LGD and PD via the observed default rates in this dataset. 
Table \ref{table:DescrStats} contains descriptive statistics about annual LGD and PD data (both AR and SG) for the whole time window.

\begin{figure}[h!]
\centering
	\includegraphics[width=9cm]{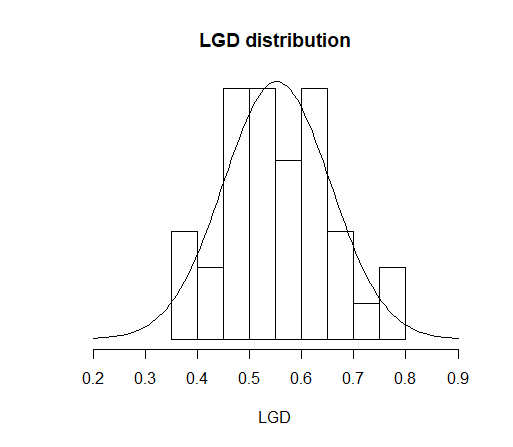}
	\caption{\small Density for $LGD$ in the dataset and for a normal distribution with same mean and std.}
\label{fig:LGDpdf}
\end{figure}

As already mentioned in the introduction, we analyse the distribution for $k = \Phi^{-1}(PD)$ and $LGD$.
The empirical distribution of LGD is shown in Figure \ref{fig:LGDpdf}. It looks well described by a normal distribution; 
moreover, modeling LGD with a Gaussian rv, the probability to observe ``non-financial"
LGD values --either negative or greater than $1$-- appears negligible.

\smallskip

\begin{figure}[h!]
\begin{tabular}{cc}
\multicolumn{2}{c}{\includegraphics[width=90mm]{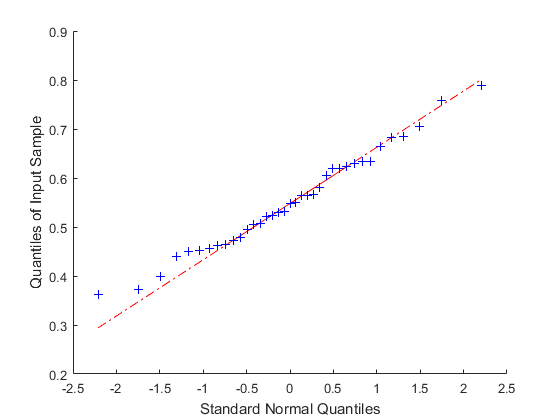} }\\[2mm]
\multicolumn{2}{c}{(a) LGD}  \\[6pt]
  \includegraphics[width=90mm]{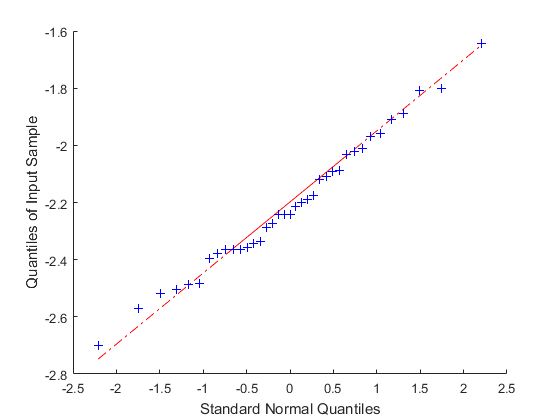} &   \includegraphics[width=90mm]{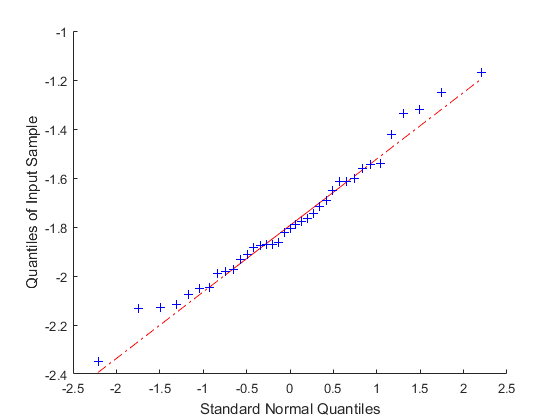} \\[2mm]
(b) $k_{AR}$ & (c) $k_{SG}$
\end{tabular}
\caption{\small Quantile-Quantile (Q-Q) plot for (a)  $LGD$, (b)  $k_{AR}$ and (c) $k_{SG}$.}
\label{fig:Quantiles}
\end{figure}

The Gaussian property of a sample can be verified in several ways: the simplest is probably the Quantile-Quantile plot.
The Quantile-Quantile plot of each parameter is shown in Figure \ref{fig:Quantiles}.
Because all the three plots tend to be close to a straight line, it seems that
the p.d.f.  of each parameter follows tightly a normal distribution.

It is also possible to verify from a quantitative perspective the normality hypothesis  via a statistical procedure. 
The Shapiro-Wilk test allows to determine if the null hypothesis of univariate
normality is a reasonable assumption regarding the
population distribution of a random sample \citep[see, e.g.][]{ShapiroWilk, Royston1982}.
Moreover, \citet{Royston1983} has extended the Shapiro-Wilk hypothesis test to the bivariate case
to verify composite normality.

The results for the  Shapiro-Wilk test statistic $W$ and the p-value are reported in Table \ref{table:SW}.
\begin{table}[h!]                                             
		\centering     
	\begin{tabular}{r c c}
		\toprule
		& $W$ & $p$-$value$ \\
		\toprule
		 $LGD$ & 0.983 & 0.840 \\
		$k_{AR}$ & 0.987 & 0.941 \\
		$k_{SG}$ & 0.979	& 0.706 \\
		 composite $LGD$-$k_{AR}$ & 0.973 & 0.509 \\
		 composite $LGD$-$k_{SG}$ & 0.984 & 0.856 \\
		\bottomrule
	\end{tabular}
	\caption{\small Shapiro-Wilk test outcome on $LGD$ and default point $k$; $W$ is the Shapiro-Wilk test statistics.
We never reject the null hypothesis of normality.}
		\label{table:SW}                                  
\end{table} 
We do not reject the null hypothesis of normality with a 10\% threshold. Notice that all p-values are above 50\%. 
Thus, we can consider normal the marginal distribution of each parameter; 
furthermore, the couple $LGD$-$k$ follows a bivariate normal distribution in both cases.

Parameter estimators $\hat{k}$ and $\sigma_k$ are reported in Table  \ref{table:MeanStd}.
The value of $\hat{k}$ is obtained from the corresponding $\hat{PD}$ via equation \eqref{eq:kHat};
in both cases (AR and SG) it is identical to the sample mean up to the third decimal digit.


\begin{table}[h!]                                             
		\centering     
	\begin{tabular}{r c c}
		\toprule
		& $AR$ & $SG$ \\
		\toprule
		 $\hat{k}$ &  -2.208 & -1.778 \\
		$\sigma_k$ & 0.237 & 0.268 \\
		\bottomrule
	\end{tabular}
	\caption{\small Parameter estimators $\hat{k}$ and $\sigma_k$ in the All Ratings and in the Speculative Grade cases. 
	The values for $\hat{LGD}$ and $\sigma_{LGD}$ are reported in Table \ref{table:DescrStats}.}
		\label{table:MeanStd}                                  
\end{table} 

\begin{figure}[h!]
\centering
	\includegraphics[width=9cm]{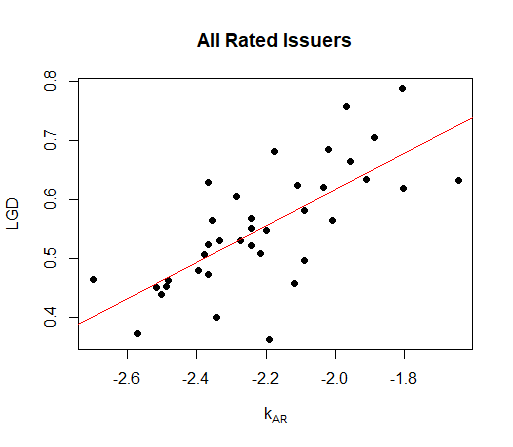}
	\caption{\small Scatter-plot of $LGD$-$k_{AR}$ and a linear regression that fits the data (with an Adjusted R-squared of $0.500$). 
We observe that the two parameters are positively correlated.}
\label{fig:ScatterAll}
\end{figure}

We can easily verify whether $LGD$ and $k$ are correlated.
In Figure \ref{fig:ScatterAll}  we show the scatter-plot of the couples $LGD$-$k$ and a linear regression that fits the data considering All Ratings. 
We can reject the uncorrelated hypothesis with a p-value $6.12 \cdot 10^{-07}$. 
The scatter plot in the Speculative Grade case looks similar; the uncorrelated hypothesis is rejected with a p-value $8.85 \cdot 10^{-05}$ in this case.

\begin{table}  [h!]                                             
		\centering     
	\begin{tabular}{r c c}
		\toprule
		& $\rho_{LGD{\text-}k}$ & $CI$ \\
		\toprule
		 All Ratings ($AR$) & 0.717 & (0.511, 0.844) \\
		Speculative Grade ($SG$) & 0.599 & (0.342, 0.773) \\
		\bottomrule
	\end{tabular}
	\caption{\small Pearson correlation between LGD and $k$ considering All Ratings and only Speculative Grade firms. 
We report the estimator $\rho_{LGD{\text-}k}$ and the $95 \%$-confidence interval (CI).}
		\label{table:PearsonCorr}                                  
\end{table} 


The estimated Pearson correlation $\rho_{LGD{\text-}k}$ between LGD and $k$ is reported (with the $95 \%$-confidence interval) in Table \ref{table:PearsonCorr}.
The value of Pearson correlation appears to be quite-high in both cases. 

\section{The measurement of model risk in
capital requirements}
\label{sec:Results}

In this section we analyse the impact on capital requirements stemming from parameter uncertainty. 

\smallskip

First, we compute the regulatory capital per unit exposure at default in the {\it na\"ive} approximation for two homogeneous portfolios, 
respectively one AR and another one SG.
The $RC$ in the {\it na\"ive} IRB approach \eqref{eqn:naiveRC}  is obtained  for a homogeneous portfolio 
considering the mean values $\hat{LGD}$ and $\hat{PD}$.
The capital requirements  in the two cases of interest are shown in Table \ref{tab:RCnaive}.
\begin{table} [h!]                                             
		\centering      
	\begin{tabular}{c c c }
		\toprule
		& $AR$ & $SG$ \\
		\midrule
		$RC^{naive}$  & 0.0866 &  0.1224 \\
		\bottomrule
	\end{tabular}
	\caption{\small Regulatory capital (RC) as a fraction of the total exposure at default in the {\it na\"ive} case for a homogeneous credit portfolio with All Ratings and 
a  credit portfolio with only Speculative Grade firms.}
\label{tab:RCnaive}
\end{table}


Then, 
the uncertainty due to the estimation of LGD and PD is taken into account measuring the {\it add-on} (\ref{eq:addon}) in different cases.
We analyse one parameter at a time (and impose the other parameter equal to its expected value) and both parameters at the same time,
considering both the  independent and the correlated case.
In this way we can ``isolate" each contribution to the {\it add-on}.

Table \ref{tab:AddOn} shows the results obtained considering either one parameter at time or the two parameters simultaneously.
When analysing the results obtained with one parameter at time,
it is interesting to observe that, in line with the literature, the largest contribution is due to the uncertainty in $PD$. 

\begin{table} [h!]                                             
		\centering
		\begin{tabular}{r l c c}
			\toprule
			 & & $AR$ & $SG$ \\
			\midrule
			$LGD$ & (only) & 5.63\% & 9.12\% \\
			$k$ & (only) & 	12.22\% & 28.87\% \\
			$LGD, k$ & (independent) & 18.67\% &	 39.54\% \\
			$LGD, k$ & (correlated) & {\bf 38.48\%} &	{\bf 65.97\%}  \\
			\bottomrule
		\end{tabular}
		\caption{\small Regulatory capital {\it add-on} due to parameter uncertainties, computed via a Monte Carlo with $N_{sim}=10^{7}$ simulations.
First, we consider the {\it add-on} due to LGD and $k$ separately, keeping the other parameter constant.
Then, on the one hand, we consider the two parameters independent and, on the other hand, correlated with the Pearson correlation $\rho_{LGD{\text-}k}$ estimated in previous section.  
The most important contribution to
the regulatory capital {\it add-on} is due to the dependency between the two parameters in both cases (AR and SG). 
The contribution of parameter uncertainty to capital requirements appears startling with an increase in the regulatory capital in the range $38 \%-66\%$.}
\label{tab:AddOn}
\end{table}

When considering the impact of both parameters,
we observe that the independent case underestimates significantly the estimation risk.
Parameter dependency, that cannot be neglected from a statistical point of view, has the most relevant impact:
it determines the most relevant contribution to the {\it add-on} in capital adequacy.\footnote{
The expected loss correction gives a small, but not negligible, contribution to the {\it add-on} 
when the two parameters are correlated,
with a correction to the {\it excess loss reserve} equal to $6 \cdot 10^{-4}$ for AR
and to  $14 \cdot 10^{-4}$ 
for SG.}

The values of {\it add-on} appear very large: 
the {\it correct} RC, that takes into account estimation noise, is significantly greater than the one computed with the {\it na\"ive} approach.
We obtain an increase in the required capital larger than $38\%$,  if All Ratings are considered, and 
almost equal to $2/3$, if we consider a credit portfolio composed only by Speculate Grade corporates.
This is the main result of this study.

\bigskip

We also perform two robustness tests. 
First, we consider a confidence level $\alpha$ equal to 
$99\%$, as it is considered in both \citet{Loffler2003} and \citet{Tarashev2010}. 
Even if the regulatory capitals with this different $\alpha$ are significantly lower, 
the {\it add-ons} look rather similar to the ones obtained with the $\alpha$ imposed by regulators for credit risk.

Second, 
we verify the impact of granularity \citep[see, e.g.][]{gordy2007}, 
i.e. we check whether we observe, for a finite number of obligors, a significant deviation from the asymptotic portfolio case. 
We have considered a small credit portfolio composed by $50$ obligors as in \citet{Loffler2003}. 
We obtain slightly higher capital requirements, but
the measure of model risk is similar to the asymptotic case. 
Both robustness tests support our empirical findings: numerical results are available upon request.


\section{Conclusions and  policy implications}
\label{sec:Conclusions}

It is common practice by risk managers to rely on a {\it na\"ive} IRB approach for capital requirement, 
where parameters are estimated with the long term averages of historical rates.
This {\it na\"ive} approximation is necessarily
a downward biased estimate of the {\it correct} regulatory capital, because  estimation noise is neglected.

\smallskip

In this letter, we show how to incorporate the inevitable uncertainty about the forecasted parameter values in measures of portfolio credit risk:
such parameter forecasting  depends on statistical hypotheses that should be tested on real datasets.
A correct quantification of capital requirements reveals that ignoring estimation noise leads to a substantial understatement of the regulatory capital;
in particular, we show that parameters' dependency plays the most relevant role in capital adequacy.

This study highlights the importance of the measure of model risk when a {\it na\"ive} approximation is implemented in credit capital requirements.
We propose to capture model risk via a succinct measure, termed
(regulatory capital) {\it add-on}, 
which is an incremental capital charge to $RC^{naive}$ for the estimation risk in IRB approaches.

\smallskip

This {\it add-on} ranges from $38\%$ (All Ratings) up to $66\%$ (Speculative Grade), where this second value could be an important benchmark for model risk.
We have shown that IRB models could be subject to significant model risk; we expect that this risk could be particularly relevant during periods of financial distress, 
which are when 
several obligors are downgraded (even to speculative grade) at the same time.
Unfortunately, these periods of financial distress are the ones when
a capital adequacy
is most needed.

\smallskip

At first glance, this result could be a cause for concern 
due to the documented degree of model risk  in credit capital requirements.

However, this result should not take us by surprise, but it should allow drawing some policy implications for capital requirements. 
For market risk, an adjustment buffer is taken into account via a a multiplication factor $m_c$ imposed by the regulators.
Regulatory capital is calculated as $m_c$ times the measured VaR, where the minimum value for $m_c$ is 3.
It has been shown that such a multiple is in line with the model risk adjustment buffer for several market risks \citep[see, e.g.][and references therein]{Boucher2014}.

Also for credit risk the Basel II accord required, if the regulator found that the regulatory capital was too low,
to apply a  multiplication factor (named {\it scaling factor}) --greater than $1$-- to the result of the credit VaR calculations, 
factor that corresponds to a --greater than $0$-- {\it add-on}
\citep[see, e.g.][p.275]{Hull_RM}.
In Basel III, the Committee has agreed to remove this {\it scaling factor}  \citep[cf.][p.6]{BaselIII}.
The main conclusion of this study from a financial policy perspective is that, to cope with the associated model risk, 
regulators should reintroduce the {\it scaling factor} at least equal to $1.4$, 
when a bank prefers to stuck with a {\it na\"ive} regulatory computation. 

\smallskip

Ultimately, this analysis has laid bare the weaknesses of the {\it na\"ive} approximation in capital requirement
and it has provided a measure of model risk on regulatory capital.
The non-negligible results observed in terms of {\it add-on} induce us to consider carefully model risk impacts on regulatory capital for banks' portfolios.
A better understanding of estimation risk in IRB approaches should lead to more robust policymaking in credit risk capital requirements.

\section*{Acknowledgments}
The authors thank all participants to the seminar at the European Investment Bank (EIB). 
We are grateful in particular to Michele Azzone, Giuseppe Bonavolont\'a, Mohamed Boukerroui, Szabolcs Gaal, Juraj Hlinicky, Aykut Ozsoy, Oleg Reichmann, 
Sergio Scandizzo and Pierre Tychon for useful comments.

The authors acknowledge EIB financial support under the EIB Institute Knowledge Programme.
The findings, interpretations and conclusions presented in this document are entirely those of the authors and should not be attributed in any manner to the EIB.
Any errors remain those of the authors.

\section*{Abbreviations}

\begin{tabular}{l l}
	\toprule
	AR & All Ratings, all corporates rated by Moody's \\ 
	ASRF & Asymptotic Single Risk Factor model \\
	cf.  & compare  (Latin: confer) \\
	CI  & $95\%$-confidence interval \\
	e.g.   & for example (Latin: exempli gratia) \\
	i.e.   & that is (Latin: id est) \\
	i.i.d. & independent identically distributed\\
	IRB    & Internal Rating Based approach \\
	p.d.f. & probability density function \\
	rv   & random variable \\
	SG & Speculative Grade corporates \\ 
	s.t.   & such that \\
	std  & standard deviation \\
	st.n.  & standard normal \\
	VaR & value-at-risk \\
	w.r.t. & with respect to \\
	\bottomrule
\end{tabular}


\section*{Notation}
\begin{tabular}{l l}
	\toprule
	\textbf{Symbol} & \textbf{Description} \\
	\midrule
    	$\varepsilon_{\bullet}$ & obligor-specific risk component, modeled as a st.n. rv \\
	$\mathbb{E}[\bullet]$ & expected value \\
	$EL^{naive}$ & expected loss $\mathbb{E}[L]$  in the {\it na\"ive} approximation \\
    $\Phi (\bullet) $ & cumulative distribution function of the st.n. rv \\
	$k$ & default point, defined as  $\Phi^{-1}(PD)$ \\
	$L$ & {\it portfolio loss rate}, i.e. total losses per unit exposure at default \\
	$LGD$ & loss-given-default \\
	$\hat{LGD}$&  mean loss-given-default \\
	$M$ & market risk variable, modeled as a st.n. rv \\
    $\mathbb{N} (\mu, \sigma^2) $ & Gaussian distribution with mean $\mu$ and variance $\sigma^2$\\
	$n$ & number of obligors in the reference portfolio \\
	$N_{sim}$ & number of simulations in the Monte Carlo method \\
	$PD$ & annual probability of default, estimated with the annual default rate in a real database\\
	$\hat{PD}$&  mean probability of default \\
	$\rho$ & obligors' asset correlation \\
	$\rho_{LGD{\text-}k}$ & Pearson correlation between LGD and $k$ \\
	$RC$ & (correct) regulatory capital \\
	$RC^{naive}$ & regulatory capital in the {\it na\"ive} approximation \\
	$\sigma^2_k$ & variance of the $k$ parameter \\
	$\sigma^2_{LGD}$ &  variance of the $LGD$ parameter  \\
	$X_{i}$ & log-asset for the $i^{th}$ obligor\\
	\bottomrule
\end{tabular}


\bigskip

\newpage
\bibliography{BiblioLetter}
\bibliographystyle{tandfx}

\end{document}